\documentclass[12pt]{article}

\usepackage{bm,graphics}

\def\putgraph#1#2{
    \vskip 0.5cm
    \centerline{\resizebox{#1}{!}{\includegraphics{#2}}}
}

\begin{document}

\title{Measurement of Mean Flows in Faraday Waves}
\author{Peilong Chen \\
   Department of Physics and Center for Complex Systems, \\
             National Central University, Chungli 320, Taiwan}

\date{\today}

\maketitle

\begin{abstract}

We measure the velocities of the mean flows which are driven by curved
rolls in a pattern formation system. Curved rolls in Faraday waves are
generated in experimental cells consisting of channels with varying
widths.  The mean flow magnitudes are found to scale linearly with
roll curvatures and squares of wave amplitudes, agreeing with
prediction from the analysis of phase dynamics expansion. The effects
of the mean flows on reducing roll curvatures are also seen.

\end{abstract}


Mean flows in pattern formation systems \cite{cross1993} are the large
scale flows whose length scales are much longer than the pattern
wavelengths.  As the patterns are convected by the mean flows, the
latters play important roles in shaping the patterns.  In fluids with
two dimensional patterns, mean flows arise when vertical vorticities are
driven by the roll curvature and amplitude modulation
\cite{siggia1981}. In this paper we present first direct measurement
of magnitudes of such mean flows and their relationship with the spatial
inhomogeneity of patterns.

In the canonical example of pattern formation, the Rayleigh-B\'enard
convection (RBC), mean flows are believed to play essential roles in
at least three kinds of patterns. The first is the appearance of
chaotic states close to onset in a large-aspect-ratio cylindrical cell
with a low Prandtl number fluid \cite{ahlers1978}. It was shown that a
key element for understanding these results is the generation of mean
flows, as modeled in the Swift-Hohenberg type model equations
\cite{greenside1988}.

In the second case, a non-Boussinesq fluid was observed to undergo a
transition from hexagonal patterns to roll states further away from
the threshold. However the roll states tend to form rotating
spiral patterns in the cylindrical cells, with different possible
numbers of arms \cite{bodenschatz1991}. Again simulations coupling
the Swift-Hohenberg equation for non-Boussinesq fluids with mean flows
were found to produce similar rotating spiral states \cite{xi1993}.

Finally, chaotic patterns consisting of many rotating spirals and
other defects were seen in RBC \cite{morris1993} and termed spiral
defect chaos (SDC). Three studies have strongly implied the importance
of mean flows. Simulations of the Swift-Hohenberg equation coupled
with mean flows have reproduced chaotic states similar to SDC
\cite{xi1993b}. Experimentally, with fluids of different Prandtl
numbers $\sigma$, it was shown that the onset of SDC increases linearly
with $\sigma$ \cite{liu1996}.  This suggests the essential role of the
mean flows whose magnitudes are decreasing with $\sigma$.  Recently a
procedure was devised to suppress mean flows in numerical simulations
of full fluid equations \cite{chiam2003}.  The results showed the
collapse of SDC to stationary textures of stripes with angular bends
when the mean flows are suppressed.

So it is well established that the mean flows are important in the
formation and dynamics of many interesting patterns in RBC.  However
direct measurement of mean flows in RBC has been difficult, due
largely to the fact that the mean flow velocity is very small.
Although the phase dynamics expansion analysis
\cite{cross1984,newell1990} has given predictions about dependence of
the mean flow magnitudes on pattern inhomogeneity, confirmation of these
predictions are lacking without direct measurement.  One experiment
has confirmed the existence of the large scale flows
\cite{croquette1986} in a distorted target pattern. However the results
were very qualitative and the information for mean flow strength is
lacking.

In this paper, we measure mean flows in a different pattern formation
system, the so called Faraday waves \cite{faradaywaves}. In a fluid
with a free surface being vibrated vertically, surface waves are
excited when the vibration amplitude exceeds a frequency-dependent
critical value.  Subharmonic standing waves usually have the lowest
threshold driving amplitudes, i.e., the wave frequency being half of
the driving frequency.  By using a suitable cell geometry, we generate
waves with curved wavefronts. Furthermore the patterns are
semi-static, namely the only pattern dynamics is the convection by
steady mean flows. Detail measurement then becomes feasible.

Singular expansion of phase dynamics in RBC
\cite{cross1984,newell1990} has predicted that near threshold the mean
flows are linear to the wavefront curvatures and the square of wave
amplitudes.  We will argue that these predictions should also be true
in Faraday waves.  We test our measured mean flow velocities with these
predictions and find good agreements.

It is noted that in Faraday waves, coupled amplitude-mean flow
equations are derived for patterns consisting of multi-mode plane
waves \cite{vegameanflow}.  The formulations predict non-vanishing
coupling between the mean flows and patterns when the patterns satisfy
some symmetry requirements.

Mean flows also appear in other non-fluid pattern formation systems,
for example in lasers \cite{lega1994}. It was shown that an asymptotic
expansion on the small detuning parameter of the Maxwell-Bloch laser
equations in the so-called class B lasers leads to the coupled
Swift-Hohenberg-mean flow equations.  In this case, the role of mean
flows is played by the density of population inversion.

We setup a typical Faraday wave system where a cell made of acrylic
plastic is mounted on a mechanical vibrator. Distilled water is used
as the working fluid and driving frequencies are in the range of 150
and 350 Hz.  An accelerometer is mounted on the cell for the
monitoring of sinusoidal oscillations.

In the range of experimental parameters we use, it is known that
the patterns appearing near threshold in an infinite system are square
patterns which consist of two sets of plane standing waves with
perpendicular wavevectors \cite{squares}.  However regular roll states
could be prepared in rectangular cells with rolls parallel to the
short sides.  For example, accurate measurement of wave amplitudes
is done using this technique \cite{wernet2001}.

It is also well-known that rolls like to approach cell boundaries
perpendicularly. This is seen most dramatically in the
Rayleigh-B\'enard convection in a circular cell \cite{cross1993}, where
rolls form two loci approximately opposite to each other on the
boundary.

\begin{figure}
\putgraph{7cm}{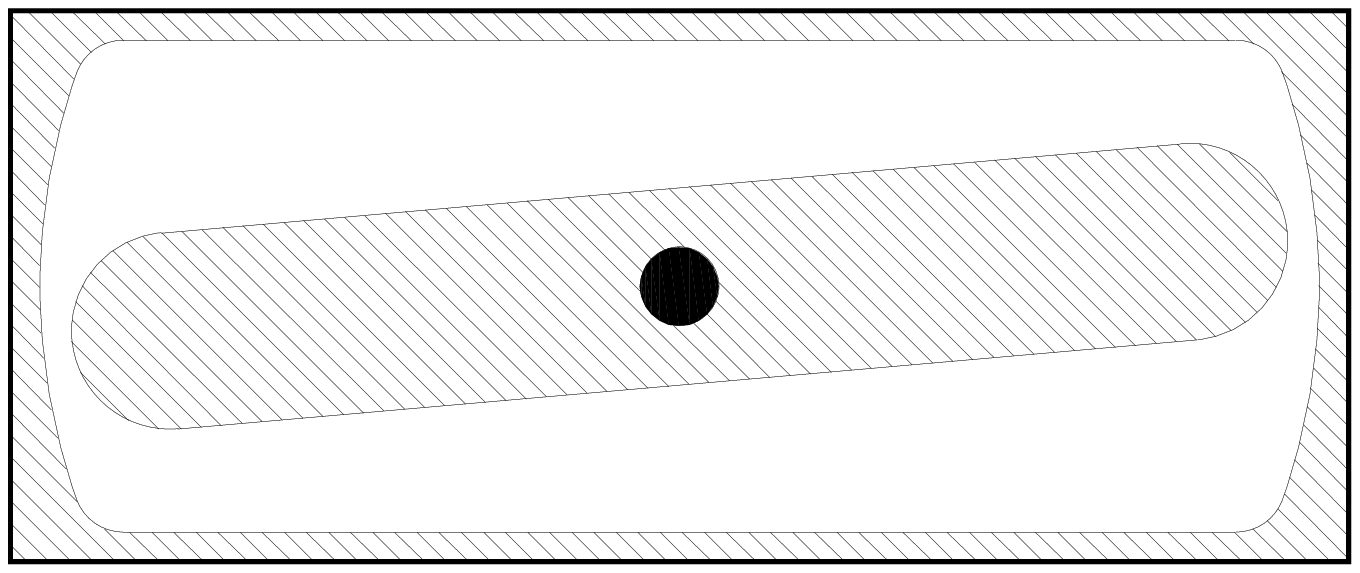}
\caption{Outline of cell A used in the experiments. 
The central divider can be rotated respect to the central axis.
The full length and width of the cell are
130mm and 50mm respectively.}
\label{fig:cella}
\end{figure}

\begin{figure}
\putgraph{7cm}{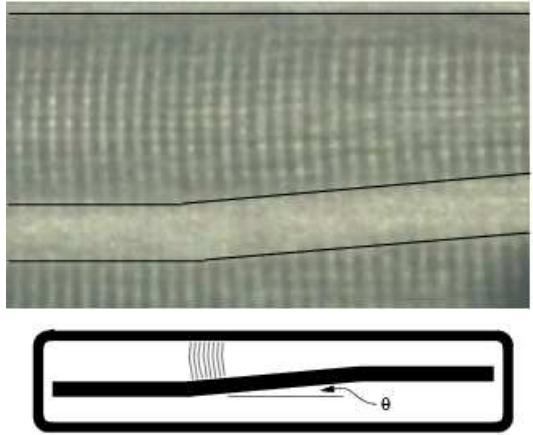}
\caption{Outline of Cell B used in the experiments. The angle $\theta$
as shown is $5^o$. The gray arcs illustrate the procedure for matching
curvature values.  The top picture shows one example of shadow graph
images for part of the cell. The full length and width of the cell are
160mm and 28mm respectively.}
\label{fig:cellb}
\end{figure}

These two properties are exploited in our experiments to produce
regular curved rolls.  Two types of cells are used as shown in
Fig.~\ref{fig:cella} and \ref{fig:cellb}. Both cells basically consist
of close-loop channels and, more importantly, the channel widths are
not uniform.  When the rolls in the converging segments try to align
perpendicularly to the non-parallel side walls, they become curved.
Furthermore, the converging segments make rolls curving toward the
same direction in terms of transversing the channel loop. Mean flows
driven by curved rolls then form circulating flows.

Cell A has a rotating central divider whose tilting angle can be
continuously adjusted.  Parallel (to the cell boundary) central
divider produces steady straight standing waves. Tilted divider causes
curved rolls which are seen to be convected by the mean flow. (Clockwise
direction in Fig.~\ref{fig:cella}.) At larger titling angles, near the
cell end with the larger opening some irregular wave patterns are
present.  Two cell B's with $\theta=5^o$ and $15^o$ are used.  Roll
patterns in cell B's are regular in the whole cell with straight rolls
in fixed-width segments.

For the data shown in this paper, cell A is filled with 15 ml of water
and driven at the frequency of 200 Hz. Cell B has 10 ml of water and
is driven at 303 Hz.

The mean flow velocity would be given by the balance between the
cell converging angles, roll curvatures and flow drags.  Flow
resistances could come from the side walls and the dissipation cost of
turning around the two ends.  It is a nice surprise to see that in
cell B the waves maintain straight rolls at the cell ends. Perhaps
this is due to the strong tendency of the roll to align perpendicular
to side walls and strong correlation between rolls.

A laser deflection method is used to measure the mean flow
velocity. A laser light is shining through the experimental cell
vertically and the sloping fluid surface deflects the light. The light
spot forming on a screen beneath the cell oscillates due
to the changing slope of the fluid surface produced by the wave
motion. Because the wave frequencies we use are much larger than the
sampling rates of both the human eyes and standard video recording
(1/30 second), a light track with a particular length appears
on the screen.

With a fixed standing wave, the length of light track depends on where
in the wave the light is shining through. If the light is exactly at
the wave peak (which becomes trough half the wave period later), the
light will not be deflected at all because the fluid surface is always
locally flat. The light track becomes a spot. On the other hand if the
light is on the nodal position of the waves, the changing slope will
be the largest and so is the light track length.

\begin{figure}
\putgraph{7cm}{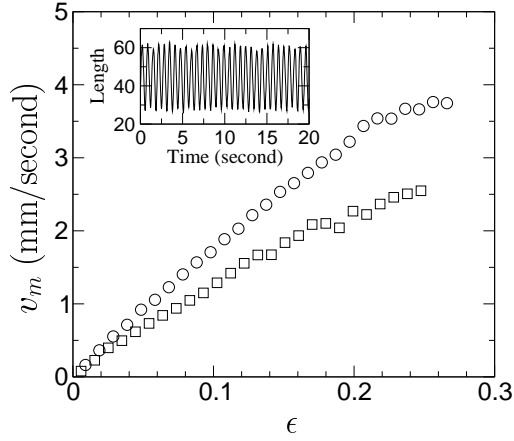}
\caption{Mean flow velocities as functions of $\epsilon$ in cell B,
         with circles for $\theta=15^o$ and squares for $\theta=5^o$.
         The inset shows an example of the light track modulation, with the
         vertical axis in an arbitrary unit.}
\label{fig:flowb}
\end{figure}

When the wave pattern is convected by the mean flow, the length of
light track on screen changes periodically as the wave is passing
through the laser light. One example is shown in the inset of
Fig.~\ref{fig:flowb}.  Since the wavelength is known, the temporal
periods of light track length modulations then correspond to the mean
flow velocities.

Images as shown in Fig.~\ref{fig:cellb} are taken by the shadow graph
method in which parallel lights shining through the transparent cell
from bottom cast an image on a screen placed horizontally on top of the
cell.  Such images are used to measure roll curvatures.


The phase dynamics expansion \cite{cross1984} predicts that near
threshold the mean flow is proportional to ${\bm
k}{\bm\nabla}\cdot({\bm k} A^2)$, with ${\bf k}$ the local wavevector
and $A$ the wave amplitude.    Physically, mean flows are driven by the large
scale (compared to the roll periodicity) pressure field $P^{(0)}_s$,
which arises from the inhomogeneity of patterns.  Here the superscript
$(0)$ denotes the zero order term on the expansion of the small
expansion parameter $\eta^2$, taken as the ratio between the
wavelength and system size.

Following Eq.~(A.26) and (A.27) of Cross and Newell \cite{cross1984},
the pressure gradient $\nabla P^{(0)}_s$ is driven by the average $\left<
\nabla\cdot [{\bm u}^{(0)}{\bm u}^{(0)}] \right>$, here ${\bm
u}^{(0)}$ is the zeroth order velocity field of rolls.  
The average is taken over one wavelength.
For
inhomogeneous roll states, large scale components then arise from
$\left< \nabla\cdot [{\bm u}^{(0)}{\bm u}^{(0)}] \right>$.  This
leads to the conclusion that the mean flow becomes $I_1 {\bm
k}\nabla\cdot({\bm k}A^2)$, with $I_1$ representing an integral over
one period of the periodic straight roll solutions.

Similar formulation in Faraday waves has not been derived.  However,
if a similar expansion analysis is carried out, we expect that the
mean flow should follow a similar form, $I_2 {\bm k}\nabla\cdot({\bm
k}A^2)$, with the proportional constant $I_2$ an
integral over a wavelength as well as a temporal wave period of the
velocity field in Faraday waves roll solutions.  The conclusions are
that the mean flow will be proportional to the roll curvature
($\sim\nabla\cdot{\bm k}$) and the square of the wave amplitude $A^2$. Our
measured flow velocity will be tested against these dependences.

A picture of roll patterns is shown in Fig.~\ref{fig:cellb}.
Direction of the mean flow is clockwise and in such direction the flow is
opposing the roll curvature. 
As the waves are seen to be convected uniformly in the whole cell,
this implies that the mean flow is uniform along the channel. This
seems contradicting the mass conservation since the channel width is
varying. However the dependence of the mean flow on the vertical
coordinate can ensure both the mass conservation and a constant
velocity near the fluid surface.

\begin{figure}
\putgraph{7cm}{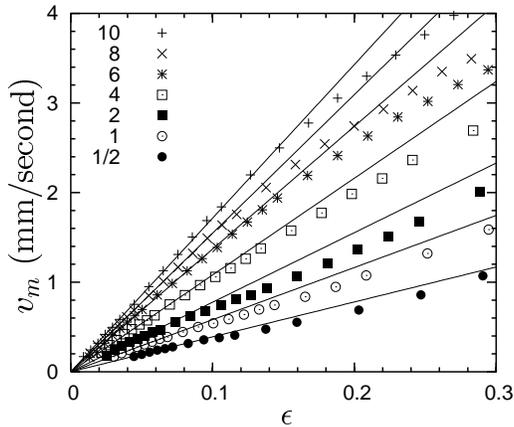}
\caption{Mean flow velocities as functions of $\epsilon$ in cell A, 
         with the numbers denoting the converging angle $\theta$.}
\label{fig:flowa}
\end{figure}

The mean flows $v_m$ measured from the convection of rolls are shown
in Fig.~\ref{fig:flowb} and Fig.~\ref{fig:flowa} as functions of the
reduced driving amplitudes $\epsilon$, defined as
$\epsilon=(f-f_c)/f_c$. Here $f_c$ is the critical driving amplitudes
and determined by fitting the data with $v_m = \alpha (f-f_c)/f_c$ at
small $\epsilon$, given $f_c$ and $\alpha$ as the fitting
parameters. The fittings are drawn as solid lines in
Fig.~\ref{fig:flowa}.  We see the mean flows $v_m$ scale linearly with
$\epsilon$, at least for small $\epsilon$.  In cell B the linear
relations extend to fairly large $\epsilon$'s, probably because the
cell B does not generate irregular wave patterns near the cell ends.
It is well established that the wave amplitudes $A$ scale linearly with
$\epsilon^{1/2}$, for example, as measured in Faraday waves
\cite{wernet2001}. Our wave amplitude data (not shown here) from the
light track length measurement also confirm this.  Thus we
first conclude that that $v_m$ is proportional to $A^2$ as predicted.

\begin{figure}
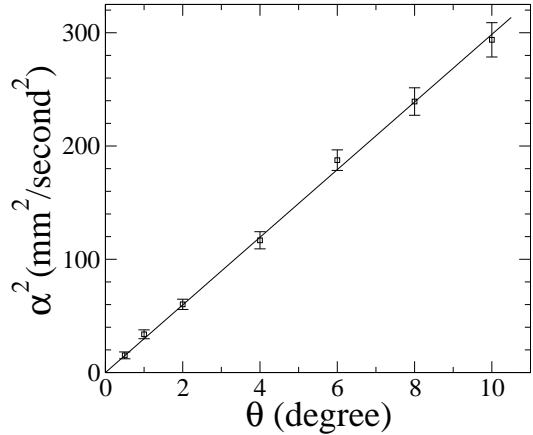

\putgraph{7cm}{alpha.eps}
\caption{Coefficient $\alpha$, as the slopes in Fig.~\ref{fig:flowa},
         plotted as a function of the converging angle $\theta$.}
\label{fig:alpha}
\end{figure}

Next we want to establish the relationship between the mean flows and
roll curvatures. In Fig.~\ref{fig:alpha} we plot from cell A the slope
$\alpha$ of the fitting between $v_m$ and $f$, with respect to the
channel converging angles $\theta$ in the experimental
cell.  The geometry of the cell limits the maximum angle at about 10
degrees. The data show a good fit with the relation
\begin{equation}
  \alpha \sim \theta^{1/2}.
  \label{eq:alphatheta}
\end{equation}

From geometrical consideration of the converging channels, if the
rolls are truely perpendicular to the side walls, the roll curvatures
$\kappa$ should be proportional to $\theta$: $\kappa\sim\theta$.
Since it is predicted that the mean flow $v_m$ is proportional to
$\kappa$, this seems contradicting the data. However we believe that the
data are also reflecting another effect, namely that the
mean flows are flowing in the direction of opposing roll
curvatures. With $\theta$ and $v_m$ have opposite effects on
curvatures, in the lowest order we might expect the dependence,
$$
  \kappa\sim\theta/v_m.
$$
Combining with $v_m\sim\kappa$ leads to $v_m\sim\kappa\sim\theta^{1/2}$.

This relation of course can be directly checked by the measurement of
roll curvatures.  However their accurate measurements from
shadow-graph images are difficult, mainly because the curved rolls are
relatively short compared to $1/\kappa$.  There are also further
complication that the curvatures are not uniform in the whole cell.
We did test the curvatures in cell B.  In cell B, since maximum
curvatures are observed to be at the beginning of the converging
segments, the following procedure is used to determine a curvature for
each image: Eight arcs of the same curvature are used to fit the
patterns, as illustrated in the bottom graph of Fig.~\ref{fig:cellb}.
Best match to the patterns gives an averaged curvature value.  For the
two B cells used, the ratios $v_m/\kappa$ are found to be (in an
arbitrary unit) $1.1\pm0.15$ for $\theta=5^o$ and $0.87\pm0.13$ for
$\theta=15^o$.  These values are consistent with a linear relation
between $v_m$ and $\kappa$. However the errors are relatively large.

In conclusion, we have measured the mean flow velocities driven by
curved rolls in Faraday waves. The dependence of the mean flow on wave
amplitudes and roll curvatures as predicted by the phase dynamics
expansion are confirmed. The effects of the mean flows on reducing
roll curvatures are also seen in our data.

This work is supported by National Science Council of Taiwan.


\end{document}